\documentclass[twocolumn,aps,amsmath,showpacs,amssymb,nofootinbib,pre]{revtex4}
\usepackage[dvips]{graphicx}
\usepackage[dvips,usenames]{color}
\usepackage{amsmath}

\begin{document}

\title{Critical dynamics on a large human Open Connectome network}

\author{G\'eza \'Odor}
\affiliation{Institute of Technical Physics and Materials Science,
Centre for Energy Research of the Hungarian Academy of Sciences} 
\address{P. O. Box 49, H-1525 Budapest, Hungary}

\date{\today}

\begin{abstract}

Extended numerical simulations of threshold models have been performed
on a human brain network with $N=836733$ connected nodes available from the 
Open Connectome project. While in case of simple threshold models
a sharp discontinuous phase transition without any critical dynamics
arises, variable thresholds models exhibit extended power-law scaling 
regions. This is attributed to fact that Griffiths effects, stemming from 
the topological/interaction heterogeneity of the network, can
become relevant if the input sensitivity of nodes is equalized.
I have studied the effects of link directness, as
well as the consequence of inhibitory connections. Non-universal power-law
avalanche size and time distributions have been found with exponents
agreeing with the values obtained in electrode experiments of the human brain.
The dynamical critical region occurs in an extended control parameter 
space without the assumption of self organized criticality.
\end{abstract}

\maketitle

\section{Introduction}

Theoretical and experimental research provides many signals for
the brain to operate in a critical state between sustained
activity and an inactive phase \cite{BP03,T10,H10,R10,Hai}. 
Critical systems exhibit optimal computational properties,
suggesting why the nervous system would benefit from such mode
\cite{LM07}. For criticality, certain control parameters need
to be tuned, leading to the obvious question why and how this
is achieved. This question is well known in statistical physics,
the theory of self-organized criticality (SOC) of homogeneous
systems has a long history since the pioneering work of \cite{Bak}. 
In case of competing fast and slow processes SOC systems can 
self-tune themselves in the neighborhood of a phase transition 
point \cite{pruessner}. 
Many simple homogeneous models have been suggested to describe 
power-laws (PL) and various other critical phenomena, very often 
without identifying SOC responsible processes.
Alternatively, it has recently been proposed that living systems 
might also self-tune to criticality as the consequence of evolution and 
adaptation \cite{adap}.

Real systems, however, are highly inhomogeneous and one must consider
if heterogeneity is weak enough to use homogeneous models to
Heterogeneity is also called disorder in 
statistical physics and can lead to such rare-region (RR) 
effects that smear the phase transitions \cite{Vojta}.
RR-s can have various effects depending on their relevancy.
They can change a discontinuous transition to a continuous one
\cite{round,round2}, or can generate so-called Griffiths Phases (GP) 
\cite{Griffiths}, or can completely smear a singular phase transition.
In case of GP-s critical-like power-law dynamics appears over an 
extended region around the critical point, causing slowly
decaying auto-correlations and burstyness \cite{burstcikk}.
This behavior was proposed to be the reason for the working 
memory in the brain \cite{Johnson}.
Furthermore, in GP the susceptibility is infinite for an entire 
range of control parameters near the critical point, providing a
high sensitivity to stimuli, beneficial for information processing.  

Therefore, studying the effects of heterogeneity is a very important 
issue in models of real system, in particular in neuroscience.  
It has been conjectured that network heterogeneity can cause GP-s 
if the topological (graph) dimension $D$, defined by
$N_r \sim r^D$ ,
where $N_r$ is the number of ($j$) nodes within topological
distance $r=d(i,j)$ from an arbitrary origin ($i$),
is finite \cite{Ma2010}. This hypothesis was pronounced 
for the Contact Process (CP) \cite{harris74}, but subsequent 
studies found numerical evidence for its validity
in case of more general spreading models \cite{BAGPcikk,wbacikk,basiscikk}.
Recently GP has been reported in synthetic brain networks 
\cite{MM,Frus,HMNcikk} with finite $D$.
At first sight this seems to exclude relevant disorder effects 
in the so called small-world network models. However, in finite systems 
PLs are observable in finite time windows for large 
random sample averages \cite{Ferr1cikk}.

Very recently we have studied the topological behavior of
large human connectome networks and found that contrary to
the small world network coefficients they exhibit topological
dimension slightly above $D=3$ \cite{CCcikk}. This is suggests
weak long-range connections, in addition to the $D=3$ dimensional 
embedding ones and warrants to see heterogeneity effects in
dynamical models defined on them. These graphs contain link weight
connection data, thus one can study the combined effect
of topological and interaction disorder, assuming a quasi static 
network.

This work provides a numerical analysis based on huge data sets of the 
Open Connectome project (OCP) \cite{OCP}, obtained by Diffusion Tensor 
Imaging (DTI) \cite{DTI} to describe {\it structural brain connectivity.}
Earlier studies of the structural networks were much smaller sized,
for example the one obtained by Sporns and collaborators, using
diffusion imaging techniques \cite{29,30}, consists of a highly 
coarse-grained mapping of anatomical connections in the human brain,
comprising $N = 998$ brain areas and the fiber tract densities between them.
The graph used here comprises $N=848848$ nodes, allowing one to run
extensive dynamical simulations on present days CPU/GPU clusters
that can provide strong evidences for possible scaling behavior.
It is essential to consider large connectomes, based on real
experimental data, even if they are coarse grained and
suffer from systematic errors and artifacts, because
synthetic networks always rely on some subjective
assumptions of the topologies.
Smaller systems near a singularity point of a phase transition,
where the correlations may diverge suffer from finite size
corrections, that can hide hide criticality or rare region effects.

\section{Models and methods}

Currently, connectomes can be estimated in humans at 1 $mm^3$ 
scale, using a combination of diffusion weighted magnetic resonance
imaging, functional magnetic resonance imaging and structural
magnetic resonance imaging scans.
The large graph "KKI-18" used here is generated by the MIGRAINE method
as described in \cite{MIG}.
Note that OCP graphs are symmetric, weighted networks, obtained
by image processing techniques, where the weights measure the
number of fiber tracts between nodes.
The investigated graph exhibits a single giant component of size 
$N = 836733$ nodes (out of $N = 848848$) and several small 
sub-components, ignored here, to disregard an unrealistic 
brain network scenario.  This graph has $8304786$ undirected edges, 
but to make it more realistic first I studied a diluted one, in which
$20\%$ of the edges made directed by a random connection removal process.
This directionality value is in between $5/126$ of \cite{unidir1} 
and $33\%$ reported in \cite{unidir2}.
I shall discuss the relevancy of this edge asymmetry assumption.

\begin{figure}[h]
\includegraphics[height=5.5cm]{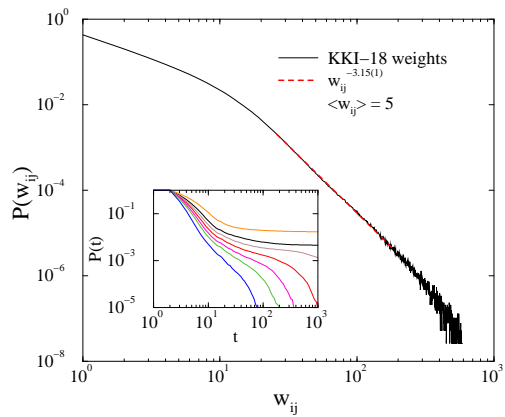}
\caption{\label{pAw} Link weight PDF of the KKI-18 OCP
graph. Dashed line: a PL fit for intermediate $w_{ij}$-s. 
Inset: Survival probability in the $k=6$ threshold model near
the transition point for
$\lambda=0.003$, $\nu=0.3$,$0.4$,$0.45$,$0.5$,$0.55$,$0.6$,$0.7$ 
(top to bottom curves).}
\end{figure}

Weights between nodes $i$ and $j$ of this graph vary between $1$ and $854$ 
and the probability density function is shown on Fig.~\ref{pAw}.
Following a sharp drop one can observe a PL region for
$20 < w_{ij} < 200$ with cutoff at large weights. The average weight 
of the links is $\simeq 5$. Note, that the average 
degree of this graph is $\langle k \rangle=156$ \cite{CCcikk}, 
while the average of the sum of the incoming weights of nodes is 
$\langle W_i\rangle = 1 / N \sum_i\sum_j w_{ij} = 448$. 

A two-state ($x_i = 0 \ {\rm or} \ 1$) dynamical spreading model was used 
to describe the propagation, branching and annihilation of activity on 
the network. 
This threshold model is similar to those of Refs.~\cite{KH,Hai}.
The dynamical process is started by activating a randomly selected node.
At each network update every ($i$) node is visited and tested if the
sum of incoming weights ($w_{i,j}$) of active neighbors reaches a given 
threshold value
\begin{equation}
\sum_{j} x_j w_{i,j}  > K \ .
\end{equation}
If this condition is satisfied a node activation is attempted
with probability $\lambda$. Alternatively, an active node is 
deactivated with probability $\nu$.
New states of the nodes are overwritten only after a full network
update, i.e. a synchronous update is performed at discrete time
steps. The updating process continues as long as there are active 
sites or up to a maximum time limit $t = 10^5$ Monte Carlo sweeps (MCs).
In case the system is fallen to inactivity the actual time step is
stored in order to calculate the survival probability $P(t)$
of runs. The average activity: $\rho(t) = 1/N \sum_{i=1}^N x_i$ and the number
of activated nodes during the avalanche $s = \sum_{i=1}^N \sum_{t=1}^T x_i$
of duration $T$ is calculated at the end of the simulations.
This stochastic cellular automaton type of updating is not expected
to affect the dynamical scaling behavior \cite{HMNcikk} and provides a
possibility for network-wise parallel algorithms. 
Measurements on $10^6$ to $10^7$ independent runs, started
from randomly selected, active initial sites were averaged over 
at each control parameter value.

By varying the control parameters, $K$, $\lambda$ and $\nu$ I
attempted to find a critical point between an active and an absorbing
steady state. At a critical transition point the survival probability 
is expected to scale asymptotically as
\begin{equation}\label{Pscal}
P(t) \propto t^{-\delta} \ ,
\end{equation}
where $\delta$ is the survival probability exponent \cite{GrasTor}.
This can be related to the avalanche
(total number of active sites during the spreading experiment)
duration scaling:
$p(t) \propto t^{-\tau_t} $ ,
via the relation $\tau_t=1+\delta$ \cite{MAval}.
In seed experiments the number of active sites initially grows as
\begin{equation}\label{Nscal}
N(t) \propto t^{\eta} \ ,
\end{equation}
with the exponent $\eta$, related to the avalanche size distribution
$p(s) \propto s^{-\tau} $ ,
via the scaling law 
\begin{equation}\label{tau-del}
\tau=(1+\eta+2\delta)/(1+\eta+\delta)
\end{equation}
\cite{MAval}.
To see corrections to scaling I also determined the local slopes
of the dynamical exponents $\delta$ and $\eta$ as the discretized, 
logarithmic derivative of (\ref{Pscal}) and (\ref{Nscal}).
The effective exponent of $\delta$ is measured as
\begin{equation}  \label{deff}
\delta_\mathrm{eff}(t) = -\frac {\ln P(t) - \ln P(t') } {\ln(t) - \ln(t')} \ ,
\end{equation}
using $t - t'=8$ and similarly can one define $\eta_\mathrm{eff}(t)$.
This difference selection has been found to be optimal in noise
reduction versus effective exponent range \cite{rmp}.

As the OCP graph is very inhomogeneous it appears that for a given set
of control parameters only the hub nodes can be activated and the
weakly coupled ones do not play any role. This is rather unrealistic
and is against the local sustained activity requirement for
the brain \cite{KH}. Indeed there is some evidence that neurons have 
certain adaptation to their input excitation levels \cite{neuroadap} 
and can be modeled by variable thresholds \cite{thres}. This adaptation
leading to some homeostasis can be assumed to be valid on the 
coarse grained, node level too.
To model nodes with variable thresholds of equalized sensitivity, 
I modified the incoming weights by normalizing them as 
$w'_{i,j} = w_{i,j}/\sum_{j \in {\rm neighb. of} \ i} w_{i,j}$. 
Although the update rules are unchanged I shall call such simulations 
"variable threshold model" studies.

\section{Dynamical simulation results}

First I summarize results for related homogeneous system. 
It is well known that in branching and annihilating models with multi-particle 
($A$) reactions $m A \to (m+k) A$, $nA \to (n-l)A$ for $m>n$ the phase 
transition in the high dimensional, mean-field limit is first order type
\cite{tripcikk}. Considering the sum of occupied neighbors as the incoming 
activation potential, there is a sudden change in the balance of 
activation/deactivation possibilities as we approach the absorbing phase,
since annihilation can occur unconditionally.
Therefore, it can be expected that for threshold models with $K > 1$, 
near and above the the upper critical dimension, which is expected to 
be $d_c\le 4$, we observe discontinuous transitions. 
First I run the threshold model on an unweighted $3$-dimensional lattice
with $N=10^6$ nodes and periodic boundary conditions. I tested the low:  
$K=2,3,6$ threshold cases, being the most possible candidates for 
the occurrence of PL dynamics. 
However, for high branching probability ($\lambda=1$), where an 
efficient neural network should work, an exponentially fast evolution 
to the inactive state occurred for any $\nu > 0.001$.
On the other hand for $\nu\to 0$ the survival probability remains
finite, but the transition is very sharp, in agreement with
the results of \cite{tripcikk} and we cannot find PL dynamics. 

In case of various quenched heterogenities a similar model, the 
CP has recently been studied on 3-dimensional lattices with diluted 
disorder in \cite{3dirft}. Extensive computer simulations gave numerical
evidence for nonuniversal power laws typical of a Griffiths phase
as well as activated scaling at criticality.
The disorder, generated by the addition of long-range connections 
has also been found to be relevant in case of CP and threshold models 
in one \cite{Ma2010} and two dimensions \cite{HMNcikk}, provided 
the probability of connections fell quickly enough i. e.: 
$p \propto r^{-s}$, with $s \ge 2D$. 
This means that in case of a $D=3$ systems the graph dimension 
remains finite and GP may be expected if $s \ge 6$.
However, in a real connectome we cannot investigate these 
heterogenities separately.

\subsection{Threshold model}

Next, I performed simulations of the threshold model on the KKI-18 graph
with $K = 1, 2, 6$, since for larger $K$-s we don't expect criticality.
Again, temporal functions did not show PL behavior.
Instead, one can observe an exponentially fast drop of $P(t)$ to zero or
to some finite value, depending on the control parameters. Discontinuous
transition occurs at very low $\lambda$-s, as shown in the inset of 
Fig.~\ref{pAw}.
Therefore, the heterogeneity of the OCP graph is not strong enough to 
round the discontinuous transition, observed on the homogeneous lattice
unlike in case of the models in \cite{round,round2}.
It appears that hubs, with large $W_i=\sum_j w_{ij}$, determine the 
behavior of the whole system, while other nodes do not play a role. 
These hubs keep the system active or inactive, ruling out the occurrence of local
RR effects as in case of infinite dimensional networks \cite{Ferr1cikk}.

\subsection{Variable Threshold Model}

\begin{figure}[h]
\includegraphics[height=5.5cm]{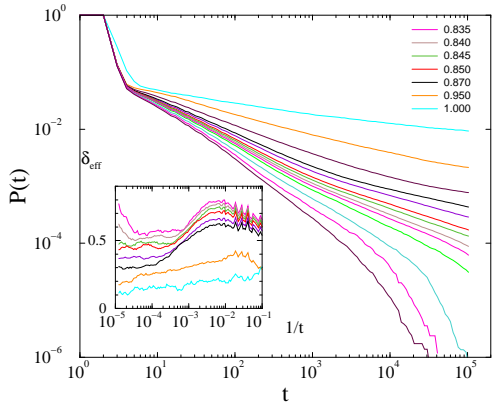}
\caption{\label{s2wlsW} Avalanche survival distribution of the
relative threshold model with $K=0.25$, for $\nu=0.95$
and $\lambda=0.8$,$0,81$,$0.82$,$0.83$,$0.835$,$0.84$,$0.845$,$0.85$,
$0.86$,$0.87$,$0.9$,$0.95$,$1$ (bottom to top curves).
Inset: Local slopes of the same from $\lambda=0.835$ to  $\lambda=1$
(top to bottom curves). Griffiths effect manifests by
slopes reaching a constant value as $1/t\to 0$.}
\end{figure}

To test this, I turned towards the simulations of variable threshold
models. The control parameter space was restricted by fixing 
$\lambda\simeq 1$, which mimics an efficient brain model.
Transitions could be found for $K < 0.5$, for higher thresholds 
the models evolve to inactive phase for any $\nu$-s.
For the time being I set $K=0.25$. 
Fig.~\ref{s2wlsW} suggests a phase transition at $\nu = 0.95$ and 
$\lambda = 0.88(2)$, above which $P(t)$ curves evolve to finite 
constant values. It is very hard to locate the transition clearly, 
since the evolution slows down and log-periodic oscillations also
add (see inset of Fig.~\ref{s2wlsW}). 
The straight lines on the log. plot of $\delta_{eff}$
at $\lambda \simeq 1$ suggest ultra slow dynamics
as in case of a strong disorder fixed point \cite{Vojta}.
Indeed, a logarithmic fitting at $\lambda = 0.88$ results in 
$P(t) \simeq \ln(t)^{-3.5(3)}$, which is rather close to the 
the $3$-dimensional strong disorder universal behavior 
\cite{3dsdrg,3dirft}.
Simulations started from fully active sites show analogous decay
curves for the density of active sites $\rho(t)$, expressing a 
rapidity reversal symmetry, characteristic of the Directed Percolation
universality class \cite{rmp}, governing the critical behavior
of such models \cite{Dickmar}. However, for the graph dimension 
of this network $D\simeq 3.2$ one should see $\delta > 0.73$ 
in case of DP universality \cite{rmp}, that
can be excluded by the present simulations.

Below the transition point for fixed $\nu=0.95$ we can find $P(t)$ 
decay curves with PL tails, characterized by the exponents $0< \delta < 0.5$,
as we vary $\lambda$ between $0.845$ and $0.88$.
In this region the avalanche size distributions also show PL
decay (Fig.~\ref{elo-t2wlsw}), modulated by some oscillations 
due to the modular network structure, but the exponents
of curves is around $\tau=1.26(2)$, a smaller value than
obtained by the brain experiments: $\tau\simeq 1.5$ \cite{BP03}. 

\subsection{Avalanche average shape}

\begin{figure}[ht]
\includegraphics[height=5.5cm]{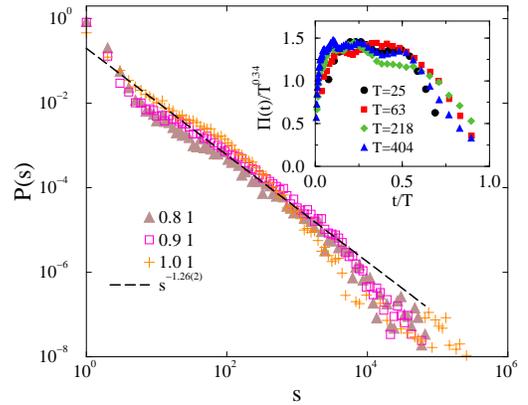}
\caption{\label{elo-t2wlsw} Avalanche size distribution of the
relative threshold model with $K=0.25$, for $\nu=1$ and
$\lambda=1,0.9,0.8$. Dashed line: PL fit to the $\lambda=0.8$ case.
Inset: Avalanche shape collapse for $T=25,63,218,404$ at
$\lambda=0.86$ and $\nu = 0.95$}
\end{figure}

I have also tested the collapse of averaged avalanche distributions
$\Pi(t)$ of fixed temporal sizes $T$ as in \cite{Fried}. The inset of
Fig.~\ref{elo-t2wlsw} shows good a collapse, obtained for avalanches 
of temporal sizes $T=25,63,218,404$ and using a vertical scaling 
$\Pi(t)/T^{0.34}$, which is near to the experimental findings  
reported is \cite{Fried}. Note, the asymmetric shapes, which
are also in agreement with the experiments and could not be
reproduced by the model of ref.~\cite{Fried}.

\subsection{Undirected links}

\begin{figure}[ht]
\includegraphics[height=5.5cm]{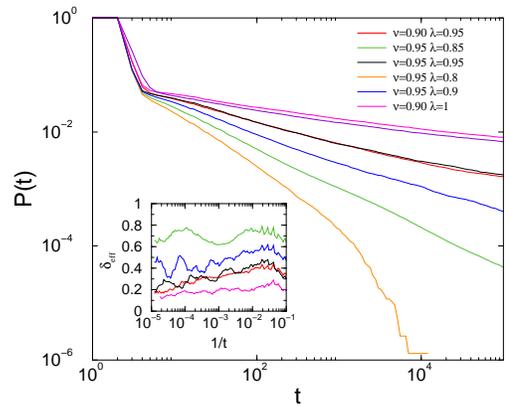}
\caption{\label{swlsW} The same as Fig.~\ref{s2wlsW} in case of
the undirected graph. Inset: Local slopes of the curves.}
\end{figure}

To test the robustness of the results simulations were also 
run on the KKI-18 network with unidirectional edges at $K=0.25$. 
Similar PL tails have been obtained as before, but for the same control 
parameters the slopes of $\ln[P(\ln{t})]$ curves were bigger, 
meaning that in the symmetric networks: $\tau_t=1.4 - 1.7$ 
(see Fig.~\ref{swlsW}) and the avalanche size distributions also 
fell down faster, characterized by $1.5 < \tau < 2$.

\subsection{Inhibitory connections}

\begin{figure}[h]
\includegraphics[height=5.5cm]{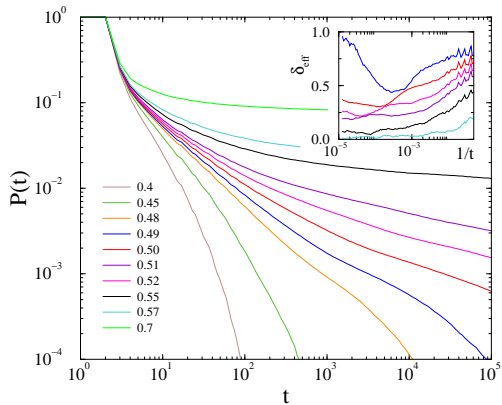}
\caption{\label{s2wlsWi3} Avalanche survival distribution of the
relative threshold model with $30\%$ inhibitory links at $K=0.1$, 
for $\lambda=0.95$ and $\nu=0.4$,$0.45$,$0.49$,$0.5$,$0.51$,$0.52$,$0.55$
$0.57$,$0.7$ (bottom to top curves). 
Inset: Local slopes of the same curves in opposite order.}
\end{figure}

\begin{figure}
\includegraphics[height=5.5cm]{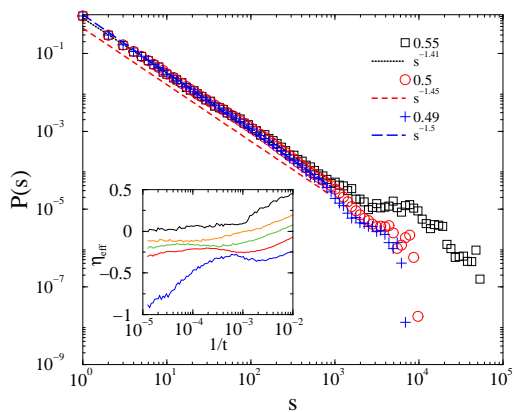}
\caption{\label{elo-t2wlsWi3} Avalanche size distribution of the
relative threshold model with $30\%$ inhibitory links at $K=0.1$, 
$\nu=0.95$ and $\lambda=0.49,0.5,0.55$. Dashed lines: PL fits.
Inset: Effective $\eta$ exponent for $\nu=0.95$ and 
$\lambda=0.49,0.5,0.51,0.51$,$0.55$ 
(bottom to top curves).}
\end{figure}

In real brain networks inhibitory connections also happen.
To model this I changed the sign of certain portion of the weights
randomly at the beginning of the simulations, 
i.e. $w'_{i,j} = -w'_{i,j}$.
This produces further heterogeneity, thus stronger RR effects. 
Figure~\ref{s2wlsWi3} shows the survival probabilities,
when $30\%$ of the links turned to inhibitory for $K=0.1$ and $\lambda=0.95$.
The critical point, above which $P(t)$ signals persistent
activity, is around $\nu=0.57$, very hard to locate clearly,
since the evolution slows down and exhibit strong (oscillating)
corrections. Below the transition point the survival exponent
changes continuously in the range $0 < \delta < 0.5$ as a response for
the variation of $\nu$ between $0.5$ and $0.57$
(inset of Fig.~\ref{s2wlsWi3}).
The corresponding avalanche size distributions (Fig.~\ref{elo-t2wlsWi3})
exhibit PL tails with the exponent $\tau\simeq 1.5$, close to the experimental value
for brain \cite{BP03}. A slight change of $\tau$ can also be observed by 
varying the control parameter below the critical point.
This variation can be seen even better on the exponent $\eta$, related 
to $\tau$ via Eq.~\ref{tau-del} (inset of Fig.~\ref{elo-t2wlsWi3}),
suggesting Griffiths effects.

For 20\% of inhibitory links the same $\tau$-s were obtained,
while 10\% of inhibition resulted in $\tau\simeq 1.3$
near the critical point.
For higher threshold values ($K=0.2,0.25$) the critical point 
shifts to smaller $\nu$ parameters but Griffiths effects are still visible. 
However, avalanche size distributions exhibit faster decay, 
characterized by:  $\tau\simeq 1.7 - 2$.

\section{Discussion and Conclusions}

Neural variability make the brain more efficient \cite{Orb16}, 
therefore one must consider its effect in modelling. 
To study this large scale dynamical simulations
have been performed on a human connectome model.
The heterogenities of an OCP graph are too weak to change 
the dynamical behavior of the threshold model of a homogeneous 
3D lattice. This seems to be true for other spreading models,
like the Contact Process \cite{Funp}.
In relative threshold models, defined by normalizing the
incoming weights, Griffiths effects have been found 
for extended control parameter spaces.
The inclusion of a $20\%$ edge directness does not affect the 
results qualitatively, reflecting insensitivity to some potential 
artifacts of DTI, like polarity detection.
Random removal of connections emulates the effect of (unknown) noise 
in the data generation and since the majority of edges is short, 
this procedure results in a relative enhancement of long connections, 
which is known to be underestimated by the DTI \cite{Tractrev}.
Scaling exponents on undirected OCPs
vary in the range $1.4 < \tau_t < 1.7$ and $1.5 < \tau < 2$,
close to neural experimental values \cite{BP03}.

Effects of other less relevant systematic errors and artifacts 
have not been investigated here.
Radial accuracy affects, for example, the end point of the tracts,
thus it influences the hierarchical levels of cortical organization.
\cite{Hilg2000}.
The present OCP exhibits hierarchical levels by construction
from the Desikan regions with (at least) two quite different scales.
Preliminary studies \cite{Funp} suggest that RR effects are enhanced 
by modularity, but not too much by the hierarchy.
Transverse accuracy determines which cortical area is connected 
to which other. However, achieving fine-grained transverse 
accuracy is difficult for DTI, not only because of limited 
spatial resolution, but also because present measures are 
noisy and indirect. We may expect that lack of precise
connection pathways are not relevant for Griffiths effects
as long as they do not affect the graph dimension for example.

The introduction of $20-30\%$ of inhibitory links, selected randomly, 
results in Griffiths effects with avalanche size and time exponents, 
which scatter around the experimental figures. 
The exponents depend slightly on the control parameters 
as the consequence of RR effects. 
Strong and oscillating corrections to scaling are also observable 
as the result of the modular structure of the connectome. 

As an earlier study \cite{CCcikk} showed certain level of 
universality in the topological features: degree distributions,
graph dimensions, clustering and small world coefficients
of the OCP graphs, one can expect the same dynamical behavior 
and Griffiths effects of these models on OCP graphs in general.
This expectation is supported further by the robustness of the
results for random changing of the network details: inhibitory 
links, directedness or loss of connections up to $20\%$. 
Therefore, one can safely take it granted that the investigated 
connectome describes well similar ones available currently
and can be considered as a useful prototype for numerical
analysis.

It is important to note that while some rough tuning
of the control parameters might be necessary to get
closer to the critical point, one can see dynamical
criticality even {\it below a phase transition point}
without external activation,
which is a safe expectation for brain systems \cite{Pris}.
Recent experiments suggest slightly sub-critical brain
states in vivo, devoid of dangerous over-activity linked to
epilepsy.

One may debate the assumption of relative thresholds, 
that was found to be necessary to see slow dynamics. 
This introduces disassortativity, enhancing RR effects
\cite{wbacikk}, besides the modularity \cite{HMNcikk}.
However, inhibitory links increase the heterogeneity 
so drastically, that a full equalization of the internal
sensitivity may not be obligatory condition for finding
Griffiths effects. This will be the target of further
studies. 

Probably the most important result of this study is 
that negative weights enable local sustained 
activity and promote strong RR effects without network 
fragmentation. Thus connectomes with high graph
dimensions can be subject to RR effects and can show
measurable Griffiths effects. Another important 
observation is that PL-s may occur in a
single network, without sample averaging, due to
the modular topological structure.

The codes and the graph used here are available on
request from the author. 

\section*{Acknowledgments}

I thank useful discussions to C.C. Hilgetag, R. Juh\'asz
and comments to M. A. Mu\~noz, M. T. Gastner.
Support from the Hungarian research fund OTKA (K109577) 
is acknowledged.

\end{document}